\documentstyle[12pt]{article}

\newcommand{\be}{\begin{equation}}
\newcommand{\ee}{\end{equation}}
\newcommand{\ba}{\begin{array}}
\newcommand{\ea}{\end{array}}
\newcommand{\bea}{\begin{eqnarray}}
\newcommand{\eea}{\end{eqnarray}}
\newcommand{\bean}{\begin{eqnarray*}}
\newcommand{\eean}{\end{eqnarray*}}

\def\up#1{\leavevmode \raise.16ex\hbox{#1}}

\def\sqr#1#2{{\vcenter{\vbox{\hrule height.#2pt
	\hbox{\vrule width.#2pt height#1pt \kern#1pt
	  \vrule width.#2pt}
	\hrule height.#2pt}}}}

\def\Journal#1#2#3#4{{#1} {\bf #2}, #3 (#4)}

\newcommand{\gapproxeq}{\lower .7ex\hbox{$\;\stackrel{\textstyle >}{\sim}\;$}}
\newcommand{\lapproxeq}{\lower .7ex\hbox{$\;\stackrel{\textstyle <}{\sim}\;$}}

\def\thebibliography#1{{\bf REFERENCES\markboth
 {REFERENCES}{REFERENCES}}\list
 {[\arabic{enumi}]}{\settowidth\labelwidth{[#1]}\leftmargin\labelwidth
 \advance\leftmargin\labelsep
 \usecounter{enumi}}
 \def\newblock{\hskip .11em plus .33em minus -.07em}
 \sloppy
 \sfcode`\.=1000\relax}

\def\NCa{{\em Nuovo Cimento} A}
\def\NCb{{\em Nuovo Cimento} B}

\def\NP{\em Nucl. Phys.}
\def\NPB{{\em Nucl. Phys.} B}
\def\PLB{{\em Phys. Lett.}  B}
\def\PRL{\em Phys. Rev. Lett.}
\def\PRD{{\em Phys. Rev.} D}
\def\ZPC{{\em Z. Phys.} C}

\begin{document}

\title{\hfill $\mbox{\small{
$\stackrel{\rm\textstyle DSF-27/99}
{\rm\textstyle hep-ph/9910447\quad\quad\quad}$}}$ \\[1truecm]
Neutrino Masses in $SO(10)$ Theories} 
\author{F. Buccella$^*$ and O. Pisanti$^*$}
\date{$~$}

\maketitle

\thispagestyle{empty}

\begin{center}
\begin{tabular}{l}
$^*$ Dipartimento di Scienze Fisiche, Universit\`a di Napoli,\\
~~Mostra d'Oltremare, Pad.19, I-80125, Napoli, Italy; \\
~~INFN, Sezione di Napoli, Napoli, Italy.\\ \\
\end{tabular}
\end{center}

\begin{abstract}
We review the status of a class of gauge unified models based on SO(10)
group and discuss the main phenomenological implications of these models in
particular for neutrino masses.
\end{abstract}

\vspace{3truecm}

\begin{tabular}{l}
\small\tt e-mail: buccella@na.infn.it  \\
\small\tt e-mail: pisanti@na.infn.it  \\
\end{tabular}

\newpage

\section{Introduction}
The standard model is a very successful theory which describes strong and
electro-weak interactions with the gauge group $G_{SM}\equiv SU(3)_c\otimes
SU(2)_L\otimes U(1)_Y$ \cite{glas} and with the left-handed fermions of
each family classified in the multiplets 
\be
(1, 1, 1) + \left( 3, 2, \frac{1}{6} \right) + \left( \bar 3, 1,
-\frac{2}{3} \right) + \left( 1, 2, -\frac{1}{2} \right) + \left( \bar 3,
1, \frac{1}{3} \right), 
\ee
and the scalar Higgses in the representation
\be
\left( 1, 2, \frac{1}{2} \right) + \left( 1, 2, -\frac{1}{2} \right). 
\ee
However, there is no explanation for the quantization of the electric
charge, $Q$, and of the weak hypercharge, $Y$, and there are too many
multiplets, 5, for each family. 

A strong indication in favour of unification with a simple larger gauge
group comes from the values of the trace of $Y$ and of the square of the
generators of $G_{SM}$ for the multiplets of one family, 
\be
\ba{ccccc}
& Tr(Y)\quad & Tr(Y^2)\quad & Tr(T_3^2)\quad & Tr(F_3^2)\quad \\[.3truecm] 
(3, 2, \frac{1}{6}) & 1 & \frac{1}{6} & \frac{3}{2} & 1 \\[.3truecm]
(\bar 3, 1, -\frac{2}{3}) & -2 & \frac{4}{3} & 0 & \frac{1}{2} \\[.3truecm]
(1, 1, 1) & 1 & 1 & 0 & 0 \\[.3truecm]
(1, 2, -\frac{1}{2}) & -1 & \frac{1}{2} & \frac{1}{2} & 0 \\[.3truecm]
(\bar 3, 1, \frac{1}{3}) & 1 & \frac{1}{3} & 0 & \frac{1}{2}
\ea
\ee
In fact, the sum on all the multiplets of $Tr(Y)$ vanishes, as is expected
for a generator of a simple group, and the sums for $Tr(T_3^2)$ and
$Tr(F_3^2)$ take the same values. Moreover, if we combine the first three
multiplets and the last two, we have $Tr(Y) = 0$ and equal values for
$Tr(T_3^2)$ and $Tr(F_3^2)$ and, in both cases, 3/5 smaller than the
contribution to $Tr(Y^2)$. All these facts strongly suggest to classify the
two sets of multiplets in the 10 and $\bar 5$ representations of $SU(5)$
\cite{gegl}, respectively. The gauge group of the standard model, $G_{SM}$,
is a maximal subgroup of $SU(5)$ and, by taking the Higgses in the adjoint
(24) and fundamental ($5 + \bar 5$) representations of $SU(5)$, it is
possible to achieve the desired breaking pattern, 
\be
SU(5) \rightarrow G_{SM} \rightarrow SU(3)_c\otimes U(1)_Y,
\ee
and the right quantization for the electric charge and the weak
hypercharge. 

Another indication concerns the coupling constants of the standard model,
which are in the relation $g_3(M_Z) > g_2(M_Z) > g_1(M_Z)$ and almost meet
at a higher scale \cite{ambf}. However, only the meeting point of $g_3$ and
$g_2$ is sufficiently high ($ > 1.6\cdot 10^{15}\, GeV$) to comply with the
experimental lower limit on proton lifetime, $\tau_{p \rightarrow e^+ +
\pi_0}^{exp} > 9\cdot 10^{32}$ years. 

Concerning the ability of $SO(10)$ in improving the predictions of minimal
$SU(5)$, it is worth to recall that there are independent motivations
\cite{geor} to consider it, rather than SU(5), as the unification gauge
group: 
\begin{itemize}   
\item
The fermions of each family can be classified in one Irreducible
Representation (IR) of $SO(10)$, the spinorial 16, with a $SU(5)$ content
of $10 + \bar 5 + 1$. In this respect, the singlet can be identified as a
left-handed antineutrino. 
\item
The accidental cancellation between the opposite anomalies of 10 and $\bar
5$ representations of $SU(5)$ is a general property of $SO(10)$ group,
which depends by the absence of a third order Casimir operator. 
\item
$SO(10)$ contains $SU(5)$ as well $SO(6)\otimes SO(4) \sim
SU(4)_{PS}\otimes SU(2)_L\otimes SU(2)_R$, first introduced by Pati and
Salam \cite{pasa}, which is very elegant in classifying the left-handed
fermions of a family in the $(4, 2, 1) + (\bar 4, 2, 1)$ representation, in
agreement with the hadron-lepton universality of the charged weak current. 
\end{itemize}

Going back to the problem of unification of the standard model constants,
to prevent conflict with experiment the evolution of $g_1$ should be
modified in such a way to cross the meeting point of $g_2$ and $g_3$. In
$SO(10)$ unified theories, where $Y = T_{3R} + (B-L)/2$ ($B$ and $L$ are
the baryonic and leptonic numbers, respectively), this may be easily
achieved considering an intermediate symmetry group $G'$ larger than
$G_{SM}$ and containing $SU(2)_R$ and/or $SU(4)$. 

\section{The spontaneous symmetry breaking of $SO(10)$}
An intermediate symmetry between $SO(10)$ and $G_{SM}$ is generally
expected, since the smallest IR's of $SO(10)$ (with the only exception of
the spino-vector 144 \cite{pisa}) have $G_{SM}$-singlets with a symmetry
larger than $G_{SM}$. One therefore expects as intermediate symmetry the
little group of the Higgs with the highest Vacuum Expectation Value (VEV).
The spinor (16) and the bispinor (126) representations have
$G_{SM}$-singlets invariant under $SU(5)$. Besides these, one is able to
identify the four interesting cases of Table~\ref{tab:break}, with the
intermediate symmetry $G' = G"\otimes SU(2)_L\otimes SU(2)_R$ \cite{aabp}.
In Table~\ref{tab:break} $D$ is a discrete symmetry exchanging the
left-handed and right-handed $SU(2)$'s \cite{kush} and we adopt the
following convention for the 10 representation of $SO(10)$: the indices
1...6 correspond to $SO(6) \sim SU(4)$ and 7...0 to $SO(4) \sim
SU(2)\otimes SU(2)$. 

\begin{table}[t]
\caption{In the breaking of $SO(10)$, the following intermediate symmetry
groups lead to phenomenologically interesting
predictions.\label{tab:break}} 
\begin{center}
\footnotesize
\begin{tabular}{|c|c|c|}
\hline
G" & \raisebox{0pt}[13pt][7pt]{\rm Higgs direction} & {\rm Representation}
\\ 
\hline
$SU(4)_{PS}\times D$ & \raisebox{0pt}[13pt][7pt]{$2~ (S_{11} + ... +
S_{66}) - 3~ (S_{77} + ... + S_{00})$} & 54 \\ 
\hline
$SU(4)_{PS}$ & \raisebox{0pt}[13pt][7pt]{$\Phi_T \equiv \Phi_{7890}$} & 210
\\ 
\hline
$SU(3)_c\times U(1)_{B-L}\times D$ & \raisebox{0pt}[13pt][7pt]{$\Phi_L
\equiv \frac{\Phi_{1234} + \Phi_{1256} + \Phi_{3456}}{\sqrt{3}}$} & 210 \\ 
\hline
$SU(3)_c \otimes U(1)_{B-L}$ & \raisebox{0pt}[13pt][7pt]{$\cos \theta~
\Phi_L + \sin \theta~ \Phi_T$} & 210 \\ 
\hline
\end{tabular}
\end{center}
\end{table}

With a model where one of the Higgses just described takes the highest VEV
and the 126 representation breaks $G'$ into $G_{SM}$ (the 16 would give too
small Majorana masses for the right-handed neutrinos \cite{witt}), one can
predict the scale of $SO(10)$ breaking, $M_X$, and the one of $G'$
breaking, $M_R$, in terms of the values of the gauge couplings at the scale
$M_Z$, for which we take $\sin^2 \theta_W (M_Z) = 0.2315\pm 0.0002$,
$\alpha_s (M_Z) = 0.120\pm 0.005$, $\frac{1}{\alpha} (M_Z) = 127.9\pm
0.09$. In the analysis we assume the {\it extended survival hypothesis}
\cite{bamo}, which states that the Higgs scalars acquire their masses at
the highest possible scale whenever this is not forbidden by symmetries,
and find the values shown in Table~\ref{tab:mrmx}. 

\begin{table}[t]
\caption{Values of the unification and intermediate scales, $M_X$ and 
$M_R$, for the four patterns of breaking of $SO(10)$ considered in the 
text.\label{tab:mrmx}} 
\begin{center}
\footnotesize
\begin{tabular}{|c|c|c|}
\hline
$G"$ & \raisebox{0pt}[13pt][7pt]{$\frac{M_X}{10^{15}\, GeV}$} &
$\frac{M_R}{10^{11}\, GeV}$ \\ 
\hline
$SU(4)_{PS}\times D$ & \raisebox{0pt}[13pt][7pt]{$0.6$} & 460 \\ 
\hline
$SU(3)_c\otimes U(1)_{B-L}\times D$ & \raisebox{0pt}[13pt][7pt]{$1.6$} &
0.7 \\ 
\hline
$SU(4)_{PS}$ & \raisebox{0pt}[13pt][7pt]{$4.7$} & 2.8 \\
\hline
$SU(3)_c\otimes U(1)_{B-L}$ & \raisebox{0pt}[13pt][7pt]{9.5} & 0.067 \\
\hline
\end{tabular}
\end{center}
\end{table}

\section{Phenomenology of SO(10) GUT's} 
In SO(10) the theoretical value of the proton lifetime is \cite{taup} 
\be
\tau_{p \rightarrow e^+ + \pi_0} = (1.1\!-\!1.4)\cdot 10^{32} \left(
\frac{M_X}{10^{15}\, GeV} \right)^4 years, 
\ee
so that the experimental lower limit, $\tau_{p \rightarrow e^+ +
\pi_0}^{exp} > 9\cdot 10^{32}$ years, excludes the intermediate symmetries
containing $D$. $M_R$ is related, {\it via} the {\it see-saw} mechanism
\cite{gera}, to the masses of the left-handed neutrinos, 
\be
m_{\nu_{iL}} = \left( \frac{m_\tau}{m_b} \right)^2 \frac{m_{u_i}^2}{M_R}
\frac{g_{2R}}{f_i}, 
\label{eq:mnu}
\ee
where $f_i$ is the Yukawa coupling of the scalars of the 126 to the $i$-th
family. 

With respect to $SU(5)$, for which proton decay is the only typical new
phenomenon, $SO(10)$ has other possible signatures, one of which is
neutron-antineutron oscillation. In fact, minimal $SU(5)$ has $B-L$ as a
global symmetry, while in $SO(10)$ this is a generator and must be
spontaneously broken, since there is no massless boson coupled to its
associated current. Very brilliant experiments have reached the lower limit
of $0.86\cdot 10^8 sec$ (90\% CL) \cite{bace} for the $n-\bar n$  time of
oscillation. Indeed, in the 126 representation there are scalars with the
proper quantum numbers to mediate $n-\bar n$ transitions. However, since
the exchange of three of them is needed, they should not be larger than
$\sim 10^{4.5}\, GeV$ to provide an oscillation time at reach of
experimental detection. The lower limit found by Baldo-Ceolin et al.
\cite{bace} allowed to prove that a longer $n-\bar n$ oscillation would
have negligible effects on the evolution of neutron stars. This happens
because the effect would also be dumped by a quantum Zeno effect, which
would make the energy loss for the oscillation and subsequent annihilation
dependent on the volume of the neutron star and not on the density
\cite{bgor}. 

\section{$SO(10)$ and neutrino masses}
A relevant difference between $SO(10)$ and $SU(5)$ concerns neutrino
masses. In $SU(5)$, like in the standard model, one does not need
right-handed neutrinos. It is possible to build Majorana masses for the
left-handed neutrinos by coupling them to $I = 1$ Higgses, which are also
not necessary in the standard model. In $SU(5)$, by classifying the
electro-weak Higgses in a $5 + \bar 5$, one gets the equality  of the mass
matrices for charged leptons and $-1/3$ quarks at the highest scale. This
prediction is modified by the renormalization group equations (RGE) into
$\frac{m_b}{m_\tau} \sim 3$ at lower scales, in qualitative agreement with
experiment. By classifying the electro-weak Higgses in the 10
representation of $SO(10)$ ($5 + \bar 5$ under $SU(5)$), one gets the
equality of the Dirac neutrino and 2/3 quark mass matrices at the highest
scale, which would be a disaster, were not for the {\it see-saw} mechanism
\cite{gera}, which transforms that prediction into the intriguing one that
neutrino masses are much smaller than the masses of the other fermions. 

Neutrino masses and mixings are now advocated to explain the anomalies in
atmospheric and solar neutrinos with square mass differences $3.5\cdot
10^{-3} eV^2$ and $2.5\cdot 10^{-5} eV^2$ (for the MSW solution
\cite{wolf}), respectively. 
\begin{table}[t]
\caption{Values of the combination $\frac{m_\tau}{m_b}~ m_t$ in
Eq.~(\ref{eq:mnu}), corresponding to the Majorana mass of right-handed
neutrinos. All the quantities are measured in GeV.\label{tab:mass}} 
\begin{center}
\footnotesize
\begin{tabular}{|c|c|c|c|}
\hline
& \raisebox{0pt}[13pt][7pt]{$M_R$} & 
$\frac{m_\tau}{m_b}~ m_t$ & $\frac{m_\tau}{m_b}~ m_c$ \\
\hline
& \raisebox{0pt}[13pt][7pt]{$2.5\cdot 10^{11}$} & 4 & 1 \\
\hline
& \raisebox{0pt}[13pt][7pt]{$2.5\cdot 10^{13}$} & 40 & 10 \\
\hline
& \raisebox{0pt}[13pt][7pt]{$2.5\cdot 10^{15}$} & 400 & 100 \\
\hline
\raisebox{0pt}[13pt][7pt]{\rm Experimental value} && $\sim 70$ & $\sim 0.5$
\\ 
\hline
\end{tabular}
\end{center}
\end{table}
This corresponds to the Dirac masses in the last two column of
Table~\ref{tab:mass}, depending on the Majorana mass of the right-handed
neutrinos (with the simplifying assumption to classify the Higgs doublets 
in the 10 representation of $SO(10)$). The orders of magnitude comply well
with the value advocated for solar neutrinos in the case of the the model
with intermediate symmetry $SU(4)_{PS}\otimes SU(2)_L\otimes SU(2)_R$
\cite{su4}, while for atmospheric neutrinos a larger Majorana mass, $\sim
10^{13}\, GeV$, would be preferred. 

It is worth reminding that the effective $\nu_L$ mass matrix is given by 
\be
M_D^T~ M_M^{-1} M_D,
\ee
and is expected to give a large $\nu_\mu-\nu_\tau$ mixing, while the quark
mixing angles in the CKM matrix \cite{ckm} are small. We have the option of
giving Majorana masses to the right-handed neutrinos either by the $\Delta
| B - L | = 2$ $SU(5)$ singlet of the 126 or by the $\Delta | B - L | = 1$
$SU(5)$ singlet of the 16. In the last case one gets only higher loop
contribution to the Majorana masses, which imply that the VEV of the 16
should be some order of magnitude larger than $M_M$. 

As we have seen, the model with $SU(4)_{PS}\otimes SU(2)_L\otimes SU(2)_R$
intermediate symmetry gives a Majorana mass for the right-handed neutrinos
in agreement with the value advocated for the MSW explanation of solar
neutrinos' anomaly. However, to give the mass requested for the atmospheric
neutrinos, one has to assume that the highest Dirac mass of the neutrinos
is $\sim 8\, GeV$; this corresponds, by evolving the mass with the RGE at
the higher scales, to a value an order of magnitude smaller than the top
mass, while the hypothesis of classifying the Higgs doublets in the 10
representation of $SO(10)$ would imply equal values for them. This is not
so unreasonable, since we know that one needs other representations to
avoid the prediction of an equal mass at the highest scale for the strange
quark and the muon and that the neutrino mixing is different from the quark
one. 

\section{SUSY $SO(10)$ models}
Besides the non-SUSY models described in the previous sections, it is worth
studying whether SUSY $SO(10)$ models can provide a higher value for the
Majorana masses of right-handed neutrinos. For a long time model builders
of SUSY gauge unified theories have been stressing that, with SUSY breaking
at $TeV$ scale, as is needed to protect the electro-weak scale, the RGE are
modified in such a way that the three gauge coupling constants meet at a
sufficiently high scale to comply with the lower limit on proton lifetime.
In that framework it is not necessary to go beyond minimal $SU(5)$ as in
non-SUSY models. Let us nevertheless study the expectations for
right-handed neutrino Majorana masses in SUSY $SO(10)$ models. 

In SUSY unified theories one has less freedom in building the Higgs
potential, which implies that it is more difficult to achieve the desired
pattern of symmetry breaking and, conversely, more meaningful the
construction of consistent models. In fact, the potential consists of two
parts, both of which non-negative, 
\bea
\sum_\alpha \left|D_\alpha\right|^2 &=& \sum_\alpha g_\alpha^2 \left|<\Phi|
Q_\alpha |\Phi>\right|^2, \nonumber \\ 
\sum_a \left|F_a\right|^2 &=& \sum_a \left|\frac{\delta F}{\delta
\Phi_a}\right|^2, 
\eea
where $Q_\alpha$ are the gauge group charges and the superpotential $F$ is
an invariant function of $\Phi$ of degree $\leq 3$. There is a
complementarity between $D_\alpha$ and $F_a$, since a necessary and
sufficient condition for a field $\Phi_a$ to give $D_\alpha = 0$ is the
existence of at least an invariant function of $\Phi_a$, $G$, such that 
\be
\left( \frac{\delta G}{\delta \Phi_a} \right)_{\Phi=<\Phi>} = k <\Phi_a^*>,
\label{eq:susycon}
\ee
with $k \neq 0$ \cite{bdfs}.

To get non-trivial zeros for the SUSY potential, one a) may exclude some
terms in $F$ (this can seem unnatural, but some discrete symmetry may
help), b) consider only invariant functions $G$ with degree $\geq 4$, or c)
(the most elegant possibility) with more than one invariant of degree $\leq
3$ obeying the condition in Eq.~(\ref{eq:susycon}). 

By studying the SUSY extensions of $SO(10)$ models previously considered,
with the Higgses responsible for the spontaneous symmetry breaking of
$SO(10)$ in the 16 + 45 or in the 126 + 54, respectively, it is easy to see
that, in both cases, one needs at least an additional 16 representation to
have vanishing $Q_\alpha$. 

\section{Conclusion}
The increasing evidence in favour of neutrino masses and mixings is a
serious hint for $SO(10)$ unification, which provides all the elements,
left-handed antineutrinos and very high Majorana masses for them, for a
successful {\it see-saw} mechanism. This is a strong encouragement for the
construction of a consistent $SO(10)$ theory. It is fair to stress that the
most convincing facts in favour of physics beyond the standard model come
from neutrino oscillations first proposed by Pontecorvo and successfully
developed by him in collaboration with Gribov and Bilenky \cite{pont}. Also
the radiochemical method to detect solar neutrinos, successfully applied in
the Homestake, Galles and Sage experiments, was invented more than fifty
years ago by him \cite{cana}.


\begin{thebibliography}{99}

\bibitem{glas}S. L. Glashow, \Journal{\NP}{22}{579}{1961}; S. Weinberg,
\Journal{\PRL}{19}{1264}{1967}; A. Salam, in {\em Elementary Particle
Theory}, ed. N. Svartholm, Almquist, and Wiksell (Stockholm, 1968); G. W.
't Hooft \Journal{\NPB}{33}{173}{1971}; \Journal{}{B35}{167}{1971}. 

\bibitem{gegl}H. Georgi and S.L. Glashow, \Journal{\PRL}{32}{438}{1974}.

\bibitem{ambf}U. Amaldi, W. de Boer, and H. Furstenau,
\Journal{\PLB}{260}{447}{1991}. 

\bibitem{geor}H. Georgi, {\em Particle and Fields}, ed. C. E. Carlson, AJP
(1975); H. Fritzsch and P. Minkowski, \Journal{Ann. Phys.}{93}{183}{1975}. 

\bibitem{pasa}J. C. Pati and A. Salam, \Journal{\PRD}{10}{275}{1974}.

\bibitem{pisa} O. Pisanti, Degree Thesis (unpublished).

\bibitem{aabp}F. Acampora, G. Amelino-Camelia, F. Buccella, O. Pisanti, L.
Rosa, and T. Tuzi, \Journal{\NCa}{108 n.3}{375}{1995}. 

\bibitem{kush}V. Kuzmin and N. Shaposhnikov, \Journal{\PLB}{92}{115}{1980}.

\bibitem{witt}E. Witten, \Journal{\PLB}{91}{81}{1980}.

\bibitem{bamo}R. Barbieri, G. Morchio, D. V. Nanopoulos, and F. Strocchi,
\Journal{\PLB}{90}{91}{1980}. 

\bibitem{taup}F. Buccella, G. Miele, L. Rosa, P. Santorelli, and T. Tuzi,
\Journal{\PLB}{233}{178}{1989}; P. Langacker, {\em Inner Space and Outer
Space}, ed. E. Kolb et al. (University of Chicago Press, 1986), p.1. 

\bibitem{gera}M. Gell-Mann, P. Ramond, and R. Slansky, in {\em
Supergravity}, (North Holland, 1980); T. Yanagida, in {\em Proceedings of
the Workshop on the Unified Theory and the Baryon Number of the Universe},
ed. O. Sawada et al., (KEK, 1979). 

\bibitem{bace}M. Baldo-Ceolin et al., \Journal{\ZPC}{63}{409}{1994}.

\bibitem{bgor}F. Buccella, C. Gualdi and M. Orlandini
\Journal{\NCb}{100}{809}{1987}. 

\bibitem{wolf}L. Wolfenstein, \Journal{\PRD}{17}{2369}{1978}; S. P.
Mikheyev and A. Yu. Smirnov, \Journal{Sov. J. Nucl. Phys.}{42}{913}{1985}. 

\bibitem{su4}D. Chang, R.N. Mohapatra, and M.K. Parida,
\Journal{\PRL}{52}{1072}{1984}; \Journal{\PRD}{30}{1052}{1984}; F.
Buccella, L. Cocco, A. Sciarrino, and T. Tuzi,
\Journal{\NPB}{274}{559}{1986}; R.N. Mohapatra and M.K. Parida,
\Journal{\PRD}{47}{264}{1993}. 

\bibitem{ckm}N. Cabibbo, \Journal{\PRL}{10}{531}{1963}; M. Kobayashi and T.
Maskawa, \Journal{Prog. Theor. Phys.}{49}{652}{1973}. 

\bibitem{bdfs}F. Buccella, J. P. Derendinger, S. Ferrara, and C.A. Savoy,
\Journal{\PLB}{115}{375}{1982}. 

\bibitem{pont}B. Pontecorvo, \Journal{Sov. Phys. JETP}{6}{429}{1958};
\Journal{}{7}{172}{1958}; \Journal{}{26}{984}{1968}; V. Gribov and B.
Pontecorvo, \Journal{\PLB}{28}{493}{1969};S.M.Bilenky and B.Pontecorvo
\Journal{\PLB}{61}{248}{1976}; Comments Nucl. and Part.Phys.7 {149}{1977}. 
 
\bibitem{cana}National Research Council of Canada, Division of Atomic
Energy, Chalk River, 1946, Report PD-205. 

\end{thebibliography}
\end{document}